\documentclass[prl,twocolumn,superscriptaddress]{revtex4-1}
\usepackage{amsmath} 
\usepackage{esint}
\usepackage{graphicx}
\usepackage{color}
\newcommand{\rev}[1]{{\color{black} {#1}}}

\begin{document}

{\bf Comment on `Fickian Non-Gaussian Diffusion in Glass-Forming Liquids'.}

A recent Letter~\cite{Rusciano2022} examined the statistics of individual particles displacements $\Delta x(t)$ over time $t$ in two-dimensional glass-formers and concluded that the corresponding probability distribution $G_s(\Delta x, t)$, called the van-Hove distribution, is non-Gaussian in a time regime where the mean-squared displacement (MSD) is Fickian, $\langle \Delta x^2(t) \rangle \propto D_s t$, where $D_s$ is the self-diffusion constant. If this analysis were correct, glass-formers would be `Fickian non-Gaussian' materials~\cite{FNG,miotto2021}. 

Here, we clarify that the multiple lengthscales and timescales reported in \cite{Rusciano2022} have either been characterized before, or are not well-defined. This leads us to dispute the conclusions that glass-formers display Fickian non-Gaussian behaviour and that this analogy fruitfully addresses the central questions regarding the nature of dynamic heterogeneity in these systems. 

Let us first recall that the main features of self-diffusion in supercooled liquids are explained by invoking two characteristic timescales~\cite{Berthier2005,Chaudhuri2007a,Chaudhuri2007b,hedges2007}. The self-diffusion coefficient $D_s$ controls the first one, $\tau_D = \sigma^2 / D_s$, where $\sigma$ is the particle size. The second one is the structural relaxation time $\tau_\alpha$ determined from usual time correlations, such as the self-intermediate scattering function. The adimensional ratio $X = \tau_\alpha/ \tau_D$ plays a special role. It controls both the amount of decoupling $X(T) \sim D_s \tau_\alpha$ (akin to violations of the Stokes-Einstein relation~\cite{Berthier2004})), and the Fickian lengthscale $\ell_F \propto \sqrt{X}$~\cite{Berthier2005}. These known results~\cite{Berthier2005,Berthier2004,Szamel2006,Chaudhuri2007a,Chaudhuri2007b,hedges2007,jorg2014} paint a picture that is inconsistent with several \rev{conclusions} reported in \cite{Rusciano2022,miotto2021} \rev{as we now show.}

Let us start with the van Hove distribution. It was found in \cite{Szamel2006} that $G_s(\Delta x,t)$ approaches a Gaussian distribution only for times much longer than $\tau_D$, a result rediscovered in \cite{Rusciano2022} with equivalent tools. However, the non-Gaussian parameter $\alpha_2(t)$ used in \cite{Rusciano2022} to reveal Gaussianity decays \rev{as a power law at large times. Hence, the gradual emergence of Gaussian behaviour from $\alpha_2(t)$ is a scale-free process and there is no characteristic timescale
after which self-diffusion is Gaussian,} although of course empirically $\alpha_2(t)$ will be small when $\langle \Delta x^2 \rangle \gg \ell_F^2$. It was found numerically~\cite{Berthier2004} and explained theoretically~\cite{Berthier2005} that it is the lengthscale $\ell_F$ that controls \rev{the} crossover in the wavevector dependence of the self-intermediate scattering function, a result \rev{ignored} in~\cite{Rusciano2022}.

For times $t < \tau_\alpha$, $G_s(\Delta x,t)$ is characterized by a Gaussian core at small $\Delta x$ and a nearly exponential tail $G_s \propto \exp(-|\Delta x|/\lambda)$ at large $\Delta x$~\cite{Chaudhuri2007a}. Refs.~\cite{Rusciano2022,miotto2021} discuss the existence and possible universality of a power law description of the time evolution of the exponential tail, $\lambda(t) \sim t^\alpha$. As noticed in \cite{Chaudhuri2007a,Chaudhuri2007b,barkai2020}, the exponential tail is generically explained by a large deviation argument, but asymptotic convergence is so slow that the actual value of $\lambda$ depends on the fitted range \rev{(see \cite{Chaudhuri2007b} for an explicit test and \cite{barkai2020} for analytic results suggesting that $\alpha=0$)}, which may explain reported discrepancies~\cite{Rusciano2022,miotto2021}. More fundamentally $\lambda(t)$ is \rev{difficult to measure, and $\alpha$ is not a novel characteristic exponent.}

Linear behaviour of the MSD is visually detected~\cite{Rusciano2022} in log-log representations after a time $\tau_D$ which \rev{grows more slowly than $\tau_\alpha$ at low temperature,} but the approach to linearity is algebraic~\cite{jeppe}. This power law approach to Fickian behaviour is again scale free and no characteristic timescale controls the emergence of Fickian behaviour in $\langle \Delta x^2(t)\rangle$; in particular $\tau_D$ does not play this role.

\rev{Even though glass-formers may appear empirically close to Fickian non-Gaussian materials, there is no characteristic timescales or lengthscales controlling the approach to either Fickian and Gaussian dynamics, and the existence of a Fickian non-Gaussian regime cannot be decided. Instead, the} salient features of self-diffusion, including algebraic approach to Fickian and Gaussian behaviours as well as nearly exponential van Hove distributions, are analytically captured by (effective) non-interacting continuous time random walk models~\cite{Berthier2005,Chaudhuri2007a,Chaudhuri2007b,hedges2007,jorg2014} based on the only two important and well-defined characteristic timescales $\tau_D$ and $\tau_\alpha$. The multiple time and length scales determined empirically in \cite{Rusciano2022,miotto2021} are either related to those, or conceptually ill-defined.

The complexity of glass-formers is that the timescales $\tau_D$ and $\tau_\alpha$ emerge from many-body interactions (disorder is self-induced) and have non-trivial temperature dependencies which are not fully understood, but from which the very rich statistics of single particle displacement naturally follows. \rev{The behaviour of supercooled liquids is very different from several Fickian non Gaussian materials, which are described by interesting, but quite different, models~\cite{prx2017}.}

We end by noting that the use of two-dimensional simulations to study the statistics of particle displacements in glass-formers is profoundly influenced \rev{at all timescales} by Mermin-Wagner fluctuations~\cite{Flenner2015,Flenner2019}, which presumably adds to the profusion of timescales reported in \cite{Rusciano2022}.

\vspace*{0.1cm}

L. Berthier$^{1,2}$, E. Flenner$^3$, G. Szamel$^3$ 

{\small $^1$Laboratoire Charles Coulomb (L2C), Universit\'e de Montpellier, CNRS, 34095 Montpellier, France 

$^2$Yusuf Hamied Department of Chemistry, University of Cambridge, Lensfield Road, Cambridge CB2 1EW, United Kingdom

$^3$Department of Chemistry, Colorado State University, Fort Collins Colorado 80523, USA}

\end{document}